\documentclass{ws-procs10x7}

\begin{document}

\newcommand{\rt}{\rightarrow}
\newcommand{\etal}{\it et al.\rm}

\newcommand{\J}{J/\psi}
\newcommand{\ppbar}{p \bar{p}}
\newcommand{\pp}{\pi^+\pi^-}
\newcommand{\ra}{\rightarrow}
\newcommand{\Jpi}{\pi^0 J/\psi}
\newcommand{\Jeta}{\eta J/\psi}
\newcommand{\Jpipi}{\pi^0\pi^0 J/\psi}
\newcommand{\gx}{\gamma\chi_{c1,c2}}
\newcommand{\ggee}{\gamma\gamma e^+e^-}
\newcommand{\ggmm}{\gamma\gamma \mu^+\mu^-}
\newcommand{\ggll}{\gamma\gamma l^+l^-}
\newcommand{\MgJ}{M_{\gamma_h,J/\psi}}
\newcommand{\Mgg}{M_{\gamma\gamma}}
\newcommand{\ptochic}{\psi(2S)\ra \gamma\chi_{c1,2}}
\newcommand{\gguu}{\gamma\gamma\mu^+\mu^-}
\newcommand{\pppp}{\psi(2S) \rt \pi^+ \pi^- J/\psi}

\newcommand{\jpsi}{J/\psi}
\newcommand{\psip}{\psi(2S)}
\newcommand{\ppp}{\pi^+\pi^-\pi^0}
\newcommand{\ar}{\rightarrow}
\newcommand{\uu}{\mu^+\mu^-}
\newcommand{\ppj}{\pi^+\pi^-J/\psi}
\newcommand{\rar}{\rightarrow}

\title{BES Results on Charmonium Decays and Transitions}

\author{Frederick A. Harris \\(for the BES Collaboration)}

\address{Department of Physics and Astronomy, The University of
  Hawaii,\\ Honolulu, Hawaii 96822,
USA\\E-mail: fah@phys.hawaii.edu}

\twocolumn[\maketitle\abstract{
Results are reported based on samples of 58 million $\jpsi$ and 14 million
$\psip$ decays obtained by the BESII experiment.  Improved branching
fraction measurements are determined, including branching fractions for
$\jpsi\to\ppp$, $\psip\ra \pi^0\J$, $\eta\J$, $\pi^0 \pi^0 J/\psi$,
anything $J/\psi$, and
$\psi(2S)\ar\gamma\chi_{c1},\gamma\chi_{c2}\ar\gamma\gamma\jpsi$.
Using 14 million $\psi(2S)$ events, $f_0(980)f_0(980)$ production in
$\chi_{c0}$ decays and $K^*(892)^0\bar K^*(892)^0$ production in
$\chi_{cJ}~(J=0,1,2)$ decays are observed for the first time, and
branching ratios are determined. 
}]

\section{Introduction}  

The Beijing Spectrometer (BES) \cite{bes1,bes2} is a general purpose solenoidal
detector at the Beijing Electron Positron
Collider (BEPC).
BEPC operates in the center of mass energy range from 2 to 5 GeV
with a luminosity at the $J/\psi$ energy of approximately
$ 5 \times 10^{30}$ cm$^{-2}$s$^{-1}$. More details on the analyses
described here can be found in the references.

\section{\boldmath $\psi(2S) \rt \gamma \gamma J/\psi$ and $X J/\psi$}

Experimental results for the processes $\psip\ra \pi^0\J$, $\eta \J$,
and $\gamma\chi_{c1,2}$ are few and date mainly from the 1970s and
80s.~\cite{PDG04} In an analysis based on a sample of
$14.0\times 10^6$ $\psip$ events collected with the BESII detector,
events of the type {$\psi(2S) \rt \gamma \gamma J/\psi$, $J/\psi \rt
e^+ e^-$ and $\mu^+ \mu^-$ are used to measure branching ratios for
$\psip\ra\pi^0\J$, $\eta\J$, and $\gamma\chi_{c1,2}$.~\cite{ggJ} 
 Fig.~\ref{fig:MJpiee} shows, after a cut
to remove the background under the
$\psip\ra\pi^0\J$ signal from $\psip\ra\gx$,
the distribution of $\gamma \gamma$ invariant mass, $\Mgg$.
The branching fractions obtained for this and the other channels are listed in
Table~\ref{results}. The BES $B(\psip\ra\pi^0\J)$
measurement has improved precision by more than a factor of two
compared with other experiments, and the $\psip\ra\eta\J$ branching
fraction is the most accurate single measurement.
\begin{figure}[htbp]
\hspace{0.2cm}
\centerline{\includegraphics[height=4.cm,width=7.5cm]{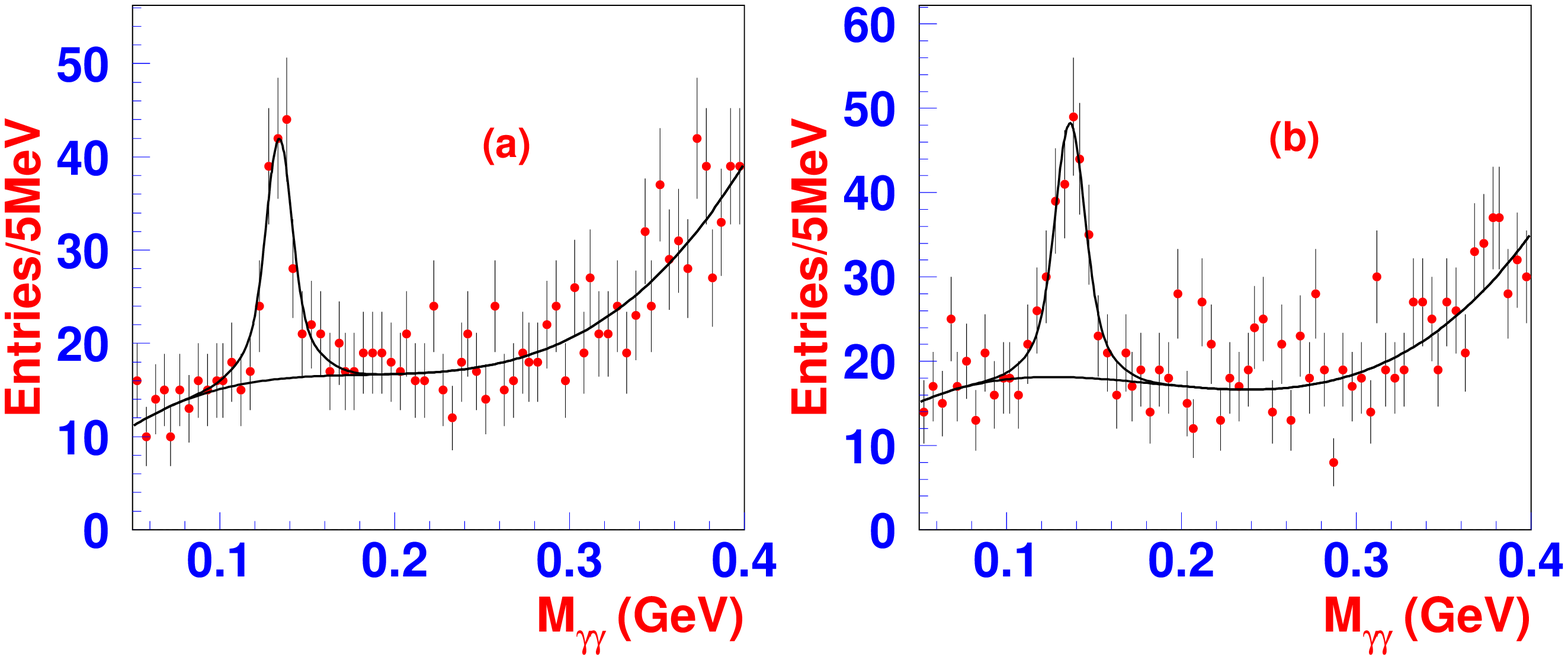}}
\caption{\label{fig:MJpiee}
  Two photon invariant mass distribution for candidate $\psip\ra\Jpi$ 
events for (a) $\ggee$ and (b) $\gguu$.}
\end{figure}
%
\begin{table*}[htbp]
\doublerulesep 0.5pt
\caption{\label{results} Results from $\psi(2S) \rt \gamma \gamma J/\psi$}
{\footnotesize
\begin{tabular}{|c|cc|cc|}              \hline 
Channel&\multicolumn{2}{c}{$\pi^0\jpsi$}&\multicolumn{2}{c|}{$\eta\jpsi$}\\\hline
Final state&$\ggee$&$\gguu$&$\ggee$&$\gguu$\\\hline
B (\%)&$0.139\pm 0.020\pm 0.012$&$0.147\pm 0.019\pm
0.013$&$ 2.91\pm 0.12\pm 0.21$&$3.06\pm 0.14\pm 0.25 $\\
Combined (\%)&\multicolumn{2}{|c|}{$0.143\pm 0.014\pm 0.012$}
&\multicolumn{2}{|c|}{$2.98\pm 0.09\pm 0.23$}\\
PDG (\%)\cite{PDG04}&\multicolumn{2}{|c|}{$0.096\pm 0.021 $}&\multicolumn{2}{|c|}
{$3.16\pm 0.22$}\\\hline\hline
Channel&\multicolumn{2}{c}{$\gamma\chi_{c1}$}&\multicolumn{2}{c|}
{$\gamma\chi_{c2}$}\\\hline
Final state&$\ggee$&$\gguu$&$\ggee$&$\gguu$\\\hline
B (\%)&$8.73\pm 0.21\pm 1.00$&$9.11\pm 0.24\pm 1.12$&
$7.90\pm 0.26\pm 0.88$&$8.12\pm 0.23\pm 0.99$\\
Combined (\%)&\multicolumn{2}{|c|}{$8.90\pm 0.16\pm
1.05$}&\multicolumn{2}{|c|}{$8.02\pm 0.17\pm 0.94$}\\
PDG (\%)\cite{PDG04}&\multicolumn{2}{|c|}{$8.4\pm 0.8$}&\multicolumn{2}{|c|}
{$6.4\pm 0.6$}\\\hline
\end{tabular}
}
\end{table*}


In another analysis, based on approximately $4 \times
10^6$ $\psip$ events obtained with the BESI detector,~\cite{bes1} a
different technique is used for measuring branching fractions for the
inclusive decay $\psip \rt {\rm anything} J/\psi $, and the exclusive
processes.~\cite{Xjpsi} Inclusive $\mu^+ \mu^-$ pairs are reconstructed,
and the inclusive branching fraction
is determined from the number of events in the
$J/\psi \rt \mu^+ \mu^-$ peak in the $\mu^+ \mu^-$ invariant mass
distribution.  The exclusive branching fractions are determined from
fits to the distribution of masses recoiling from the $J/\psi$, $m_X$, with
Monte-Carlo determined distributions for each individual channel, where
$m_X$ is
determined from energy and momentum conservation.

To separate $\psi(2S) \rt J/\psi \pi^0
\pi^0$ and $\psi(2S) \rt J/\psi \pi^+ \pi^- $ events, $m_X$ histograms
for events with (see Fig.~\ref{fig:none}) and without additional
charged tracks are fit simultaneously. 
Ratios of the studied branching fractions to that for $B(\psip \rt
\pi^+ \pi^- J/\psi)$ are reported.  This has that advantage that many
of the systematic errors largely cancel.

\begin{figure}[!htb]
\begin{center}
\centerline{\epsfysize 5 cm
\epsfbox{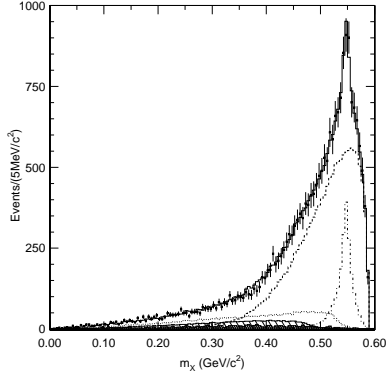}}
\caption{\label{fig:none} Fit of the $m_X$ distribution for $\psi(2S)
  \rt X J/\psi, J/\psi \rt \mu^+ \mu^-$ events with
no additional charged tracks.  Shown are the data (points with error 
bars), the component histograms, and the final fit.  For the components, 
the large, long-dash histogram is $\psi(2S) \rt J/\psi \pi \pi$, the
narrow, dash-dot histogram is $\psi(2S) \rt J/\psi \eta $, the broad,
short-dashed histogram is $\psip \rt \gamma \chi_{c1},  \chi_{c1} \rt
\gamma J/\psi$, the broad,
hatched histogram is $\psip \rt \gamma \chi_{c2},  \chi_{c2} \rt \gamma J/\psi$, and the lowest
cross-hatched histogram is the combined $e^+ e^- \rt \gamma \mu^+
\mu^-$ and $e^+ e^- \rt \psi(2S), \psi(2S) \rt (\gamma)\mu^+ \mu^-$
background. The final fit is the solid histogram.
}
\end{center}
\end{figure}

\begin{table*}[!t]
\caption{\label{tab:results} Branching ratios and branching
  fractions from $\psi(2S) \rt X J/\psi$.  PDG04-exp results are single measurements or
  averages of measurements, while PDG04-fit are results of their
  global fit to many experimental measurements. The BES results in the second
  half of the table are calculated using the PDG04 value of $B_{\pi \pi}
  = B(\psip \rt J/\psi \pi^+ \pi^- ) = (31.7 \pm 1.1) \%$. }
{\footnotesize
{\begin{tabular} {|l|c|c|c|} \hline
Case                                      &   This result &
PDG04-exp & PDG04-fit  \\ \hline
$B(J/\psi \: ({\rm anything})/B_{\pi \pi}$   & $1.867  \pm 0.026  \pm 0.055$ & $2.016
\pm 0.150$  &$1.821 \pm 0.036$   \\
$B(J/\psi\pi^0 \pi^0)/B_{\pi \pi}$ & $0.570 \pm 0.009 \pm
0.026$ & - & $0.59 \pm 0.05$ \\
$B(J/\psi \eta)/B_{\pi \pi}$ & $0.098 \pm  0.005 \pm
0.010$ & $0.091 \pm 0.021$ & $0.100 \pm 0.008$\\ 
$B(\gamma \chi_{c1})B(\chi_{c1} \rt \gamma J/\psi)/B_{\pi \pi}$ & $0.126
\pm 0.003 \pm 0.038$ & $0.085 \pm 0.021$  & $0.084 \pm 0.006$ \\ 
$B(\gamma \chi_{c2})B(\chi_{c2} \rt \gamma J/\psi)/B_{\pi \pi}$ & $0.060
\pm 0.000 \pm 0.028$ & $0.039 \pm 0.012$  & $0.041 \pm 0.003$ \\ \hline
$B(J/\psi \: ({\rm anything})$ (\%)  & $59.2  \pm 0.8 \pm 2.7  $ & $55 \pm 7$ & $57.6 \pm 2.0$ \\ 
 $B(J/\psi \pi^0 \pi^0)$ (\%)  &  $18.1 \pm 0.3 \pm 1.0 $ & -- & $18.8 \pm 1.2$\\
 $B(J/\psi \eta)$  (\%)    &  $3.11 \pm 0.17 \pm 0.31 $ & $2.9 \pm
0.5$ & $3.16 \pm 0.22$\\ 
 $B(\gamma \chi_{c1})B(\chi_{c1} \rt \gamma J/\psi)$ (\%)  & $4.0 \pm
 0.1 \pm 1.2$ & $2.66 \pm 0.16$& $2.67 \pm 0.15$\\ 
 $B(\gamma \chi_{c2})B(\chi_{c2} \rt \gamma J/\psi)$ (\%)  & $1.91 \pm 0.01 \pm 0.86$
& $1.20 \pm 0.13$& $1.30 \pm 0.08$\\ \hline
\end{tabular}}
}
\end{table*}

The final branching fraction ratios and branching fractions are shown
in Table~\ref{tab:results}.
$B(J/\psi
\:{\rm anything})/B(\pi^+ \pi^- J/\psi)$ and $B(\eta J/\psi)/B(\pi^+ \pi^- J/\psi)$ have smaller errors than
the previous results.
The agreement
with the PDG fit results is good, and $B(
 \eta J/\psi)$ agrees well with that from $\psi(2S)  \rt \gamma
\gamma J/\psi$ decays above.

\section{\boldmath 
 $\chi_{cJ}$ decays }

Using $\psi(2S) \rt \gamma \chi_{cJ}$ radiative
 decays in the
 14 million $\psi(2S)$ event sample, 
BES recently studied  $\chi_{cJ} \rt p \bar{p}$ \cite{ppbar} and $\Lambda\bar
 \Lambda$.~\cite{L}
Here, we report on the analysis of $\pi^+\pi^-\pi^+\pi^-$
final states from $\chi_{c0}$ decays, where evidence for
$f_0(980)f_0(980)$ production is obtained for
the first time,~\cite{f0f0} and
on the analysis of $\pi^+\pi^-K^+K^-$
final states from $\chi_{cJ}~(J=0,1,2)$ decays, where
signals of $\chi_{cJ}$ decays
to $K^*(892)^0\bar K^*(892)^0$
are observed for the first time.~\cite{kstarkstar}

The $\pi^+\pi^-\pi^+\pi^-$ invariant mass distribution for $\psi(2S)
\rt \gamma \pi^+\pi^-\pi^+\pi^-$ events that satisfy a four-constraint
kinematic fit is shown in Fig. 3. There are clear peaks corresponding
to the $\chi_{cJ}$ states. 
\begin{figure}[hbtp]
\begin{center}
\epsfxsize=4.5cm\epsffile{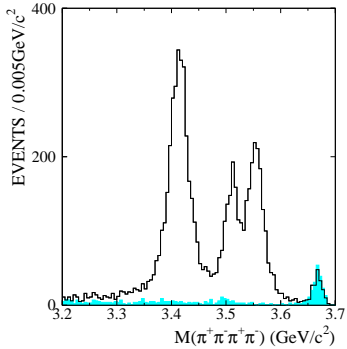}
\label{cpkp}
\caption{The $\pi^+\pi^-\pi^+\pi^-$ invariant mass spectrum for
selected  $\psi(2S) \rt \gamma \pi^+\pi^-\pi^+\pi^-$ events. The
shadow histogram shows the spectrum for Monte Carlo simulated
background events. The highest mass peak corresponds to
charged track final states that are kinematically fit with an
unassociated photon.}
\end{center}
\end{figure}
Figure 4 shows the scatter plot of $\pi^+\pi^-$ versus $\pi^+\pi^-$
invariant mass for events in the $\chi_{c0}$ peak.
A clear
$f_0(980)f_0(980)$ signal can be seen. There are
hints of $\rho^0\rho^0$ and $f_0(1370)f_0(1370)$~(or
$f_2(1270)f_2(1270)$) signals.

\begin{figure}[htbp]
\begin{center}
\epsfxsize=4.5cm\epsffile{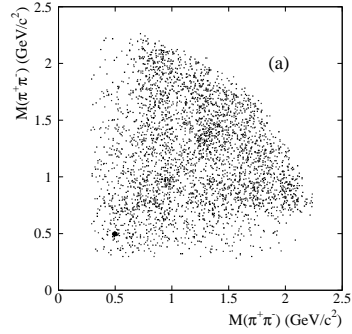}
\label{cpkp1}
\caption{Scatter plot of $\pi^+\pi^-$ versus $\pi^+\pi^-$ invariant
mass for selected $\gamma\pi^+\pi^-\pi^+\pi^-$ events with
$\pi^+\pi^-\pi^+\pi^-$ mass in the $\chi_{c0}$ peak.}
\end{center}
\end{figure}


The number of $f_0(980)f_0(980)$ events and the corresponding
background are estimated from the scatter plot.
The resulting branching fraction is
${\cal
B}(\psi(2S)\to\gamma\chi_{c0}\to \gamma f_0(980)f_0(980)\to \gamma\pi^+\pi^-\pi^+\pi^-) = (6.5\pm
1.6\pm 1.3)\times 10^{-5},$
and
using the PDG2004 value for
${\cal B}(\psi(2S)\to\gamma\chi_{c0})$  \cite{PDG04}, we obtain 
${\cal
B}(\chi_{c0}\to
f_0(980)f_0(980)\to\pi^+\pi^-\pi^+\pi^-) = (7.6\pm
1.9~(\mbox{stat})\pm 1.6~(\mbox{syst}))\times 10^{-4}.$
This may help in understanding the nature of the controversial $f_0(980)$.


A similar analysis has been done for $\psi(2S) \rt \gamma
\pi^+\pi^-K^+K^-$.
The invariant mass distribution for the $\pi^+\pi^-K^+K^-$ events
that satisfy the 4C kinematic fit shows
clear peaks corresponding to the $\chi_{cJ}$ states.
The scatter plots of $K^-\pi^+$ versus $K^+\pi^-$
invariant
masses for events 
in each of  the $\chi_{cJ}$ peaks show
clear $K^*(892)^0\bar K^*(892)^0$
signals in
all $\chi_{cJ}$ decays.
After sideband subtraction, the $K^*(892)^0\bar K^*(892)^0$ mass
spectrum is fitted with
three Breit-Wigner functions folded with Gaussian resolutions, as
shown in Fig.~5.

\begin{figure}[htbp]
\begin{center}
\epsfxsize=4.5cm\epsffile{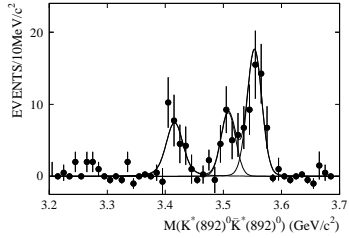}
\label{cpkp5}
\vspace*{-5pt}
\caption{The $K^*(892)^0\bar K^*(892)^0$ invariant mass spectrum
  for $\psi(2S) \rt \gamma K^*(892)^0\bar K^*(892)^0$ events
fitted with three resolution smeared Breit-Wigner functions.}
\end{center}
\end{figure}

The preliminary branching fractions are
${\cal B}(\psi(2S)\to\gamma\chi_{c0}\to\gamma K^*(892)^0\bar
K^*(892)^0) = (1.53\pm0.29\pm0.26)\times 10^{-4},$
${\cal B}(\psi(2S)\to\gamma\chi_{c1}\to\gamma K^*(892)^0\bar
K^*(892)^0) = (1.40\pm0.27\pm0.22)\times 10^{-4},$
${\cal B}(\psi(2S)\to\gamma\chi_{c2}\to\gamma K^*(892)^0\bar
K^*(892)^0) = (3.11\pm0.36\pm0.48)\times 10^{-4},$
and
with the PDG world average values of
$\psi(2S)\to\gamma\chi_{cJ}$  \cite{PDG04}, we get
${\cal B}(\chi_{c0}\to K^*(892)^0\bar
K^*(892)^0) = (1.78\pm0.34\pm0.34)\times 10^{-3},$
${\cal B}(\chi_{c1}\to K^*(892)^0\bar
K^*(892)^0) = (1.67\pm0.32\pm0.31)\times 10^{-3},$
${\cal B}(\chi_{c2}\to K^*(892)^0\bar
K^*(892)^0) = (4.86\pm0.56\pm0.88)\times 10^{-3}.$

\section{$B(J/\psi \rt \pi^+ \pi^- \pi^0$)}

The largest $J/\psi$ decay involving hadronic resonances is $J/\psi
 \rt \rho(770) \pi$.  Its branching fraction has been reported by many
 experimental
 groups~\cite{PDG04}
 assuming all $\pi^+ \pi^- \pi^0$ final states come from $ \rho(770)
 \pi$.
Here, we present two independent measurements of this branching
fraction.~\cite{3pi}.

The first, using 58 million $J/\psi$ events, is an absolute measurement of $\jpsi\ar\ppp$ directly,
where 219691 $\ppp$ candidates are selected.
The branching fraction is
$B(\jpsi\ar\ppp)=(21.84\pm0.05\pm 2.01)\times 10^{-3}$.
The Dalitz plot of $m_{\pi^+\pi^0}$ versus $m_{\pi^-\pi^0}$ is shown in
Fig. \ref{dalitz}. Three  bands are clearly visible in the plot, 
 corresponding to $\jpsi\to\rho\pi$;
$\jpsi\ar\ppp$ is
strongly dominated by $\rho\pi$.
\begin{figure}[htbp]
\centerline{\epsfig{figure=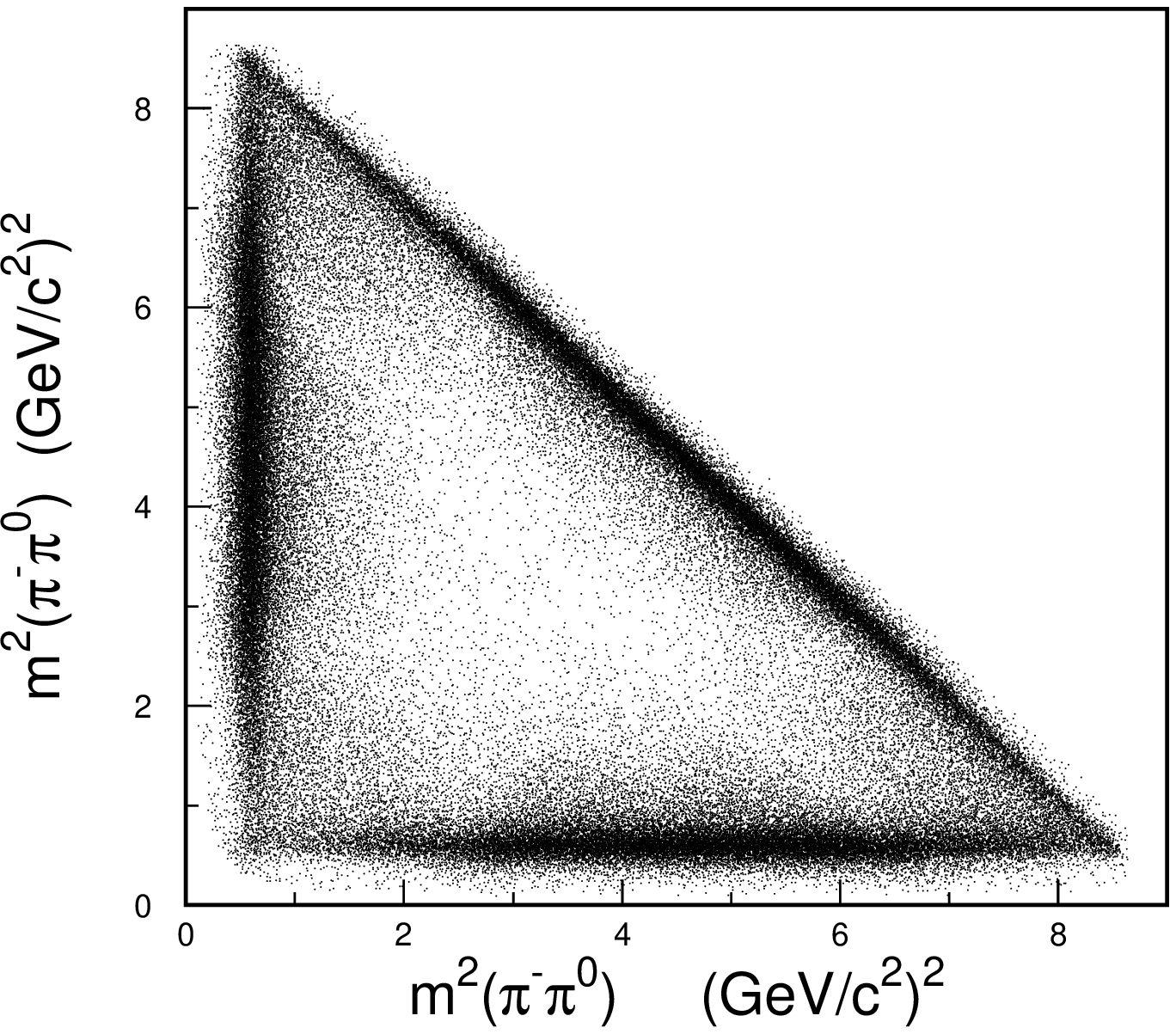,height=4.5cm}}
\caption{\label{dalitz}
  The Dalitz plot for $\jpsi\to\ppp$.}
\end{figure}

The second measurement, based on 14 million $\psi(2S)$ events,
 is a relative measurement obtained from a comparison
of the rates for
%
%
\begin{eqnarray*}
\psi(2S)\rightarrow\pi^+\pi^- &J/\psi& \\
                     &\hookrightarrow &\pi^+\pi^-\pi^0 ~~~~~~~~~~~~~~~~(I) \\
 {\rm and}                 &\hookrightarrow & \mu^+\mu^- ~~~~~~~~~~~~~~~~~(II)
\end{eqnarray*}
Here, many
systematic errors mostly cancel. 

The branching fraction obtained is
$B(J/\psi\rightarrow\pi^+\pi^-\pi^0)= (20.91\pm0.21\pm1.16)\times 10^{-3}.$
%
%
The results of the two measurements are in good agreement.
Their weighted mean is
$$B(J/\psi\rightarrow\pi^+\pi^-\pi^0)=(2.10\pm0.12)\% .$$
The result obtained is higher than those of previous
measurements and has better precision.

\section{Acknowlegments} 
I want to thank 
my BES colleagues for their efforts on the work
reported here.

\end{document}